\documentclass[twocolumn,showpacs,preprintnumbers,amsmath,amssymb,floatfix]{revtex4}%
\usepackage{graphicx}
\usepackage{amsmath}

\usepackage[center]{subfigure}
\usepackage{bigints}
\usepackage{cancel}
\usepackage{amsmath}
\usepackage{mathbbol}
\usepackage[center]{subfigure}
\usepackage{tikz}
\usepackage{standalone}
\usepackage{color}

\usepackage{mathtools}

\begin{document}
\author{Sara Najem }
\affiliation{ National Center for Remote Sensing, National Council for Scientific Research (CNRS), Riad al Soloh, 1107 2260, Beirut, Lebanon}

\begin{abstract}
We explore the relation between urban road network characteristics particularly circuitry,  street orientation entropy and the city's topography on the one hand and the building's orientation entropy on the other in order to quantify their effect on the city's solar potential.  These statistical measures of the road network reveal the interplay between the built environment's design and its sustainability. 

\end{abstract}
 \date{\today}

\title{``Along the sun-drenched roadside"\footnote{R. M. Rilke, \textit{Ahead of all parting: The selected poetry and prose of Rainer Maria Rilke} (Modern Library, 2015).}: On the interplay between urban street orientation entropy and the buildings' solar potential}

\maketitle
Light as a physical phenomenon profoundly impacted first cities formation by controlling space utility and comfort. As a symbol replete with social and mythical significance it shaped the built environment's architecture as well as the urban sensory experience \cite{shepperson2017sunlight} and its relation to the city's compactness such as buildings' density, nearest neighbor ratio, site coverage, and volume to area ratio is evidently organic \cite{shepperson2017sunlight,paz2016conceiving,Mohajeri:2016ev}. As cities grew subject to landform constraints, socio-economic factors and historic contexts \cite{Mohajeri:2014ej}, their annual solar irradiation which is the amount of solar radiation per unit area has been shown to decrease with the increase of the aforementioned compactness measures as a result of the interplay between the buildings and their neighbors' shadows \cite{Mohajeri:2016ev}. 

Further, the relation between the city and its streets pattern has been explored through the lenses of space syntax \cite{hillier1984social},  fractals \cite{bovill1996fractal,batty1994fractal} as well as statistical analysis \cite{Giacomin:2015iv,Levinson:2011gc} where  the  city is typified by its blocks shape factors which constitute its fingerprint \cite{louf2014typology}. Additionally, the correlation between the density of roads and insolation has been established \cite{Mohajeri:2014kw}. Incidentally, the streets are a product of the interplay between space and light--although not exclusively--as they grow to ensure the connectedness of the city's different components and consequently form a space filling network acting as the underlying infrastructure which sustains societal functionality. The complexity of this network can be quantified with several indicators such as circuitry, which measures the degree of the roads' divergence from Euclidean distance, length distributions, as well as the entropies of the streets' respective orientations and lengths \cite{2011PhR...499....1B,Levinson:2012gca,Gudmundsson:2013bk}, where entropy is a measure of their spread: zero entropy is indicative of a peaked distribution whereas the high level of heterogeneity is typified by a high entropic state. 

Additionally, the number of buildings has been shown to scale with the street length \cite{najem2017solar}, and subsequently the total solar potential of the city was linked to the street length distribution. However, the effect of street orientation on that of the buildings remained unexplored. In this Letter we endeavor to explore the relation between network circuitry and street orientation entropy and that of the buildings' and subsequently explore their link to the solar potential. In addition we explore the effect of cities' landforms or equivalently the distribution of their topographic elevations on their street network characteristics.  These measures can be used to evaluate the built environment's sustainability and give lead to optimal design in relation to infrastructure. 


For this purpose we retrieved the road networks and  building's footprints from OpenStreetMap \cite{osm2016} of the cities in Table \ref{my-label}. Their solar potentials $\mathcal{P}$, which is defined as the product of the buildings' yearly irradiation by their corresponding footprint areas in units of gigawatt hours per
year (GWh/year), were estimated by Mapdwell \cite{mapdwell2016} with the exception of Beirut (Lebanon) for which we used the building's footprints and elevations along with the city's topography to
evaluate $\mathcal{P}$ using the Solar Analyst
of ArcGIS \cite{najem2017solar,fu2000solar}.

\begin{table}[!htp]
\centering
\caption{$\mathcal{P}$ of multiple cities are given together with their corresponding street orientation entropy, building orientation entropy and network circuitry.}
\label{my-label}
\begin{tabular}{ccccccc}
City &  $\mathcal{P}$  & $\mathcal{S}_{\mathrm {street }}$&  $\mathcal{S}_{\mathrm {building}} $ &$\mathcal{C}$ \\ 
&     [GWh/year]& $_\mathrm{orientation}$ &$_\mathrm{orientation}$\\ \hline
Beirut (Lebanon)  & 370 & 3.32 &2.73 & 1.05\\ \hline
Boston (MA) &  1641 & 3.49& 2.75 &1.10\\ \hline
Boulder (CO) &  731 & 3.34 &2.71& 1.15\\ \hline
Cambridge (MA) & 340 &  3.51&2.69& 1.08\\ \hline
Lo Barnechea (Chile) & 582 & 3.55&2.77&1.42\\ \hline
New York (NY) & 13,330 &  3.55&2.71&1.06\\ \hline 
Portland (OR) & 8,123 & 3.23 &2.33&1.11\\ \hline
San Francisco (CA)   & 3,992  & 3.31&2.46&1.08\\ \hline
 Washington (DC) & 2,295 & 3.40 &2.67&1.11
 \end{tabular}
\end{table}

\begin{figure}[!htp]
   \includegraphics[width=0.35\textwidth]{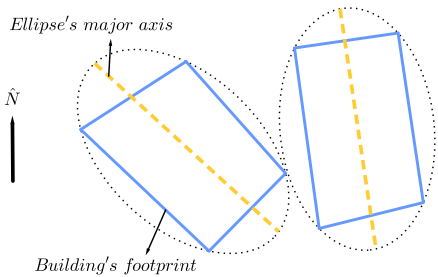}
     \caption{Two buildings footprints are shown in blue along their respective ellipses  whose major axes (in orange) define the weighted orientation. }      
\label{ellipsepoly}

\end{figure}
\begin{figure}[!htp]
  \subfigure[]{\label{rose1}
 \includegraphics[width=0.15\textwidth]{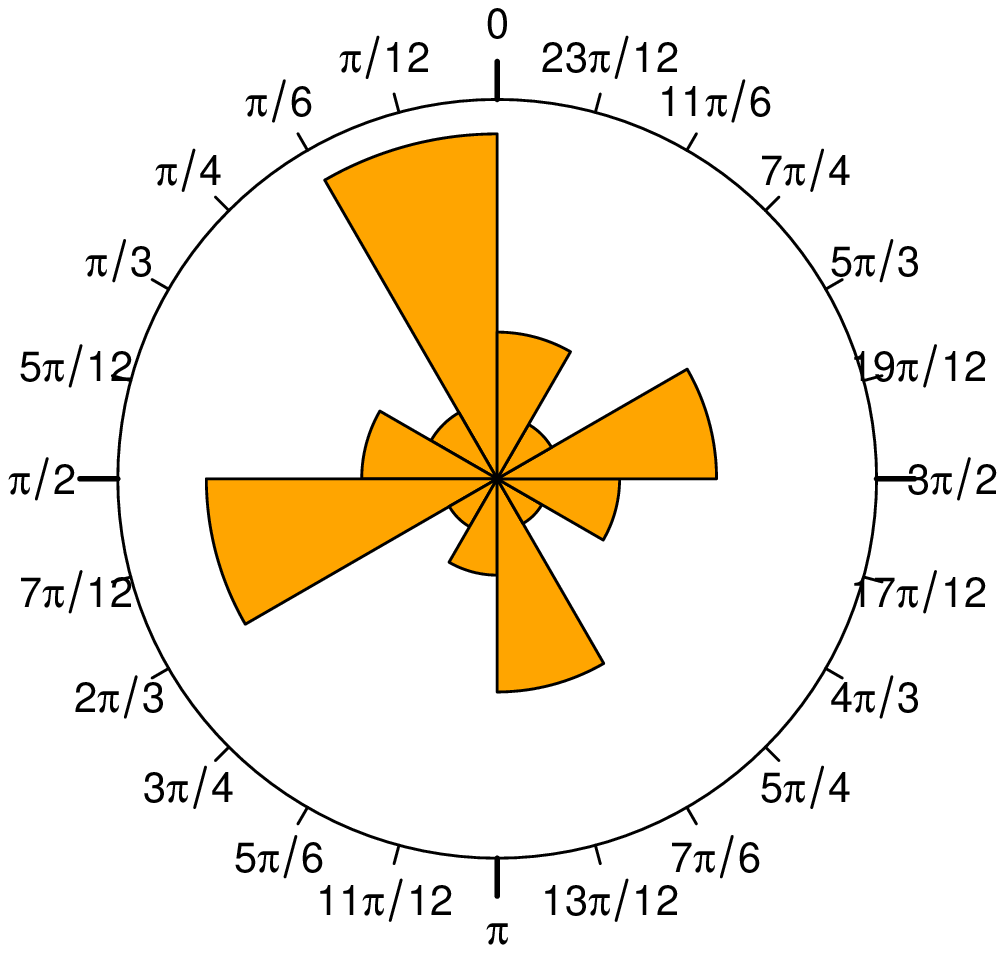}}
 \subfigure[]{\label{rose2}
   \includegraphics[width=0.15\textwidth]{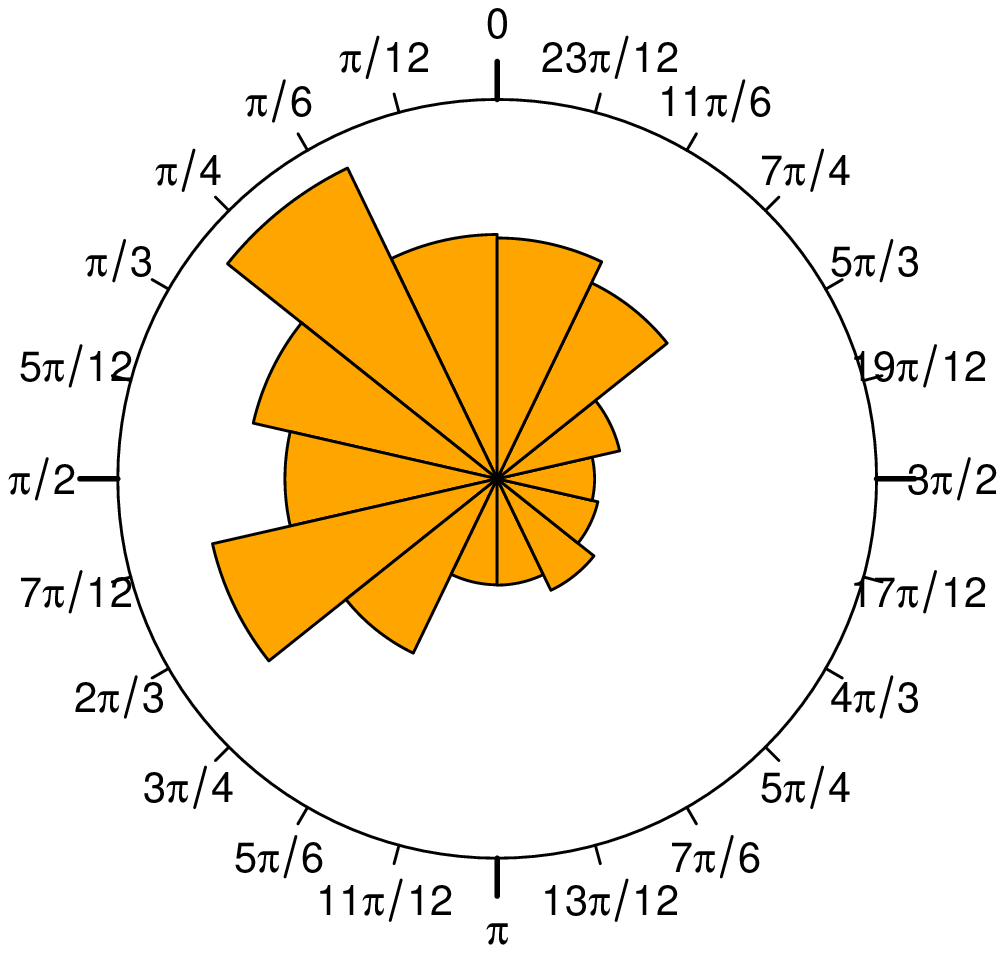}}
    \subfigure[]{\label{rose9}
   \includegraphics[width=0.15\textwidth]{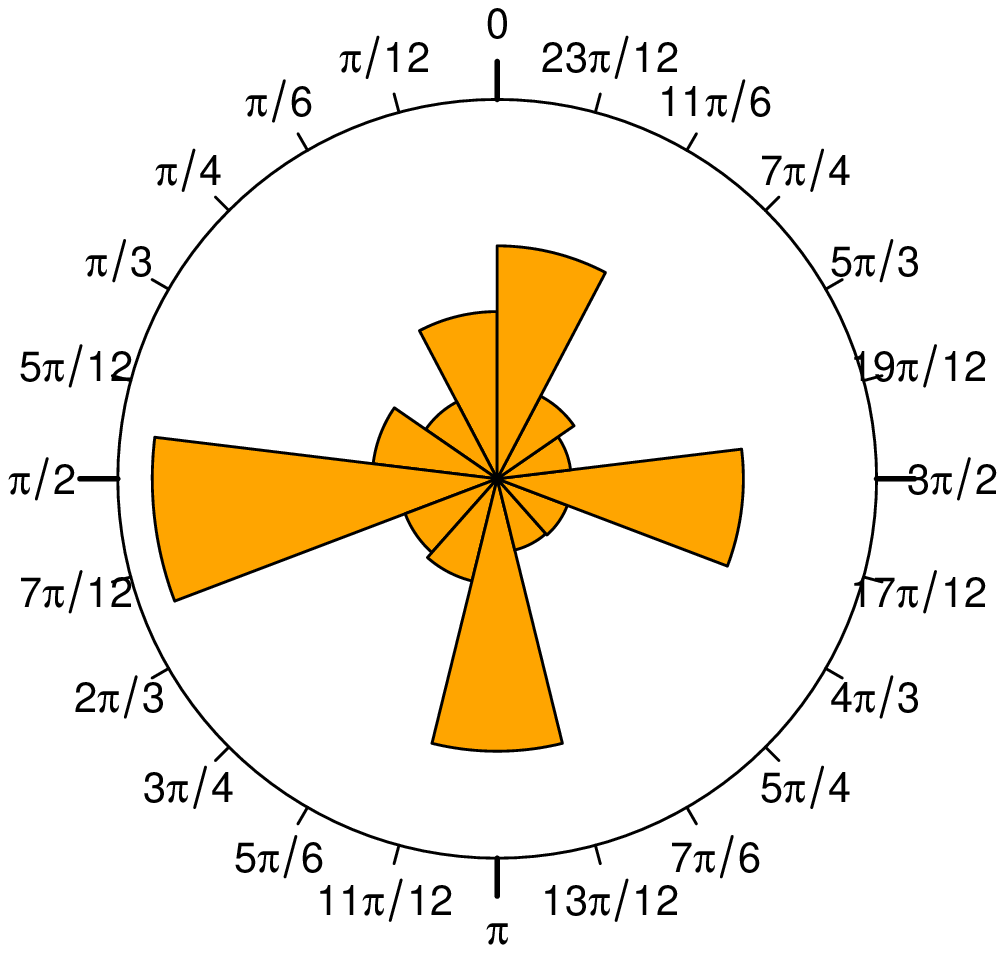}}
   \subfigure[]{\label{rose3}
   \includegraphics[width=0.15\textwidth]{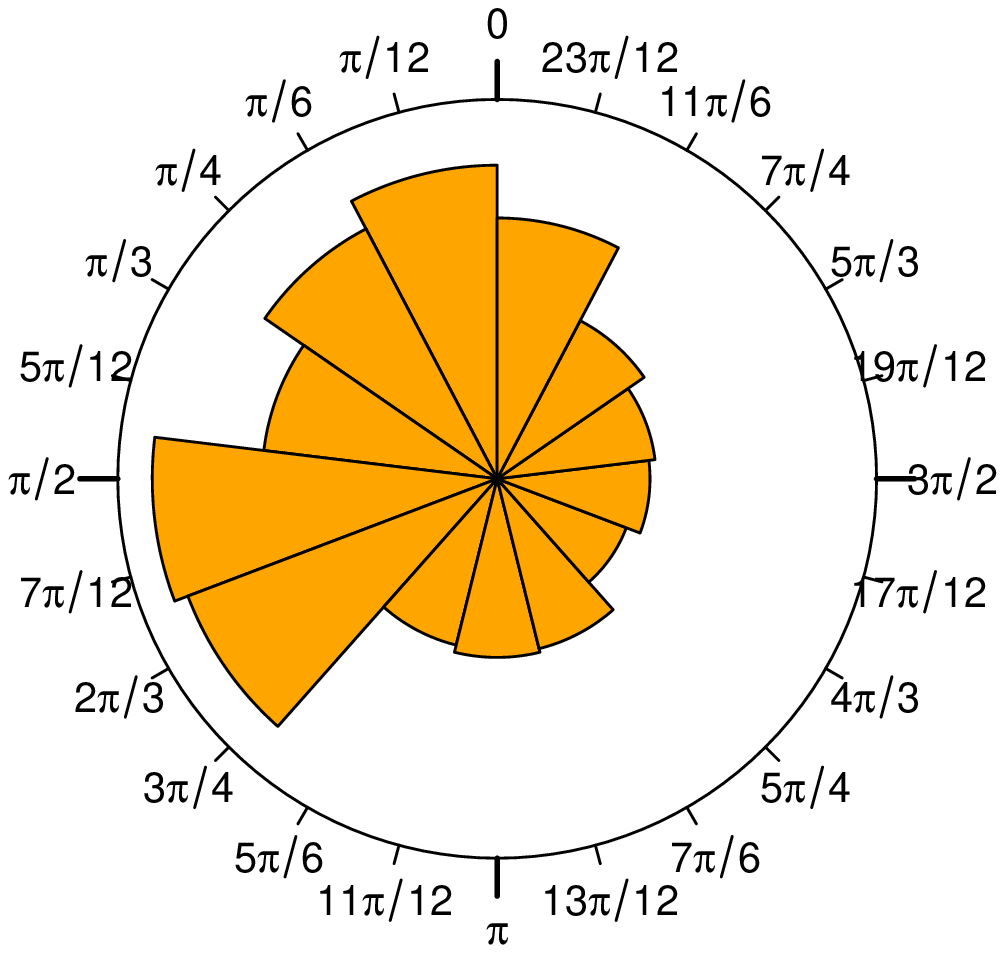}}
      \subfigure[]{\label{rose6}
   \includegraphics[width=0.15\textwidth]{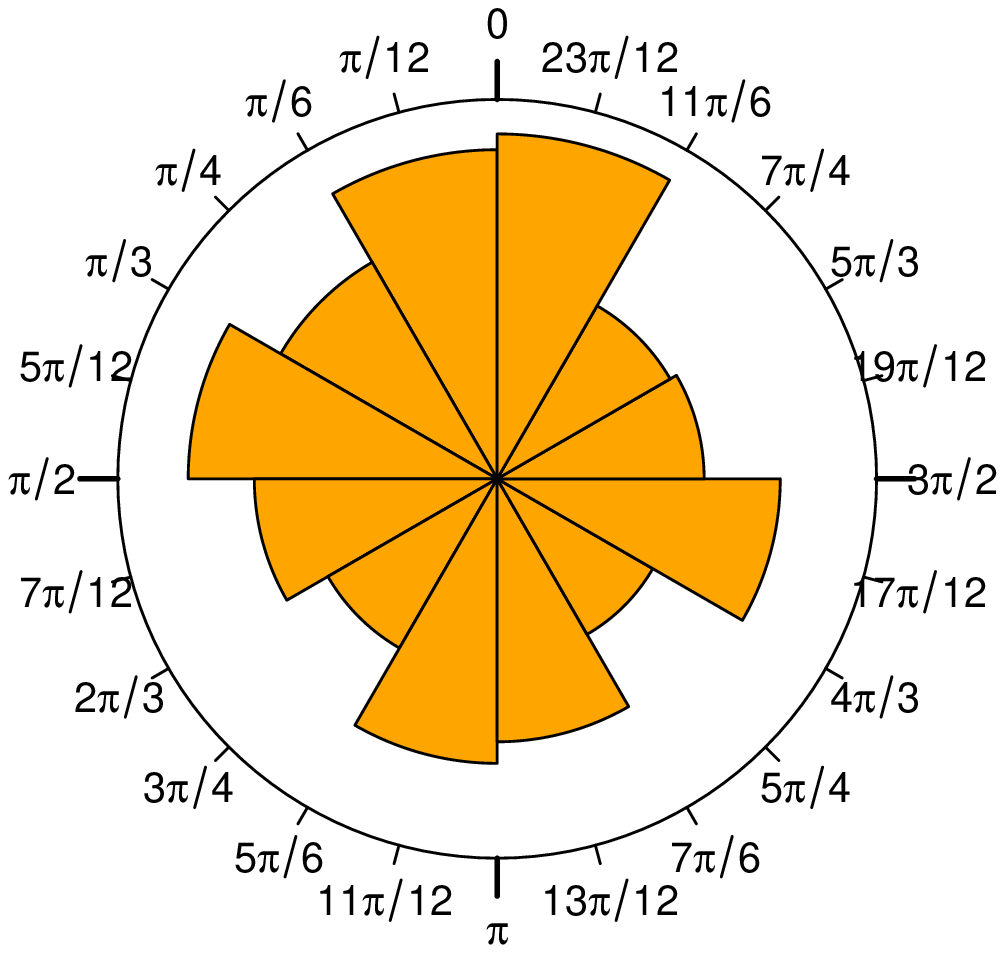}}
   \subfigure[]{\label{rose5}
   \includegraphics[width=0.15\textwidth]{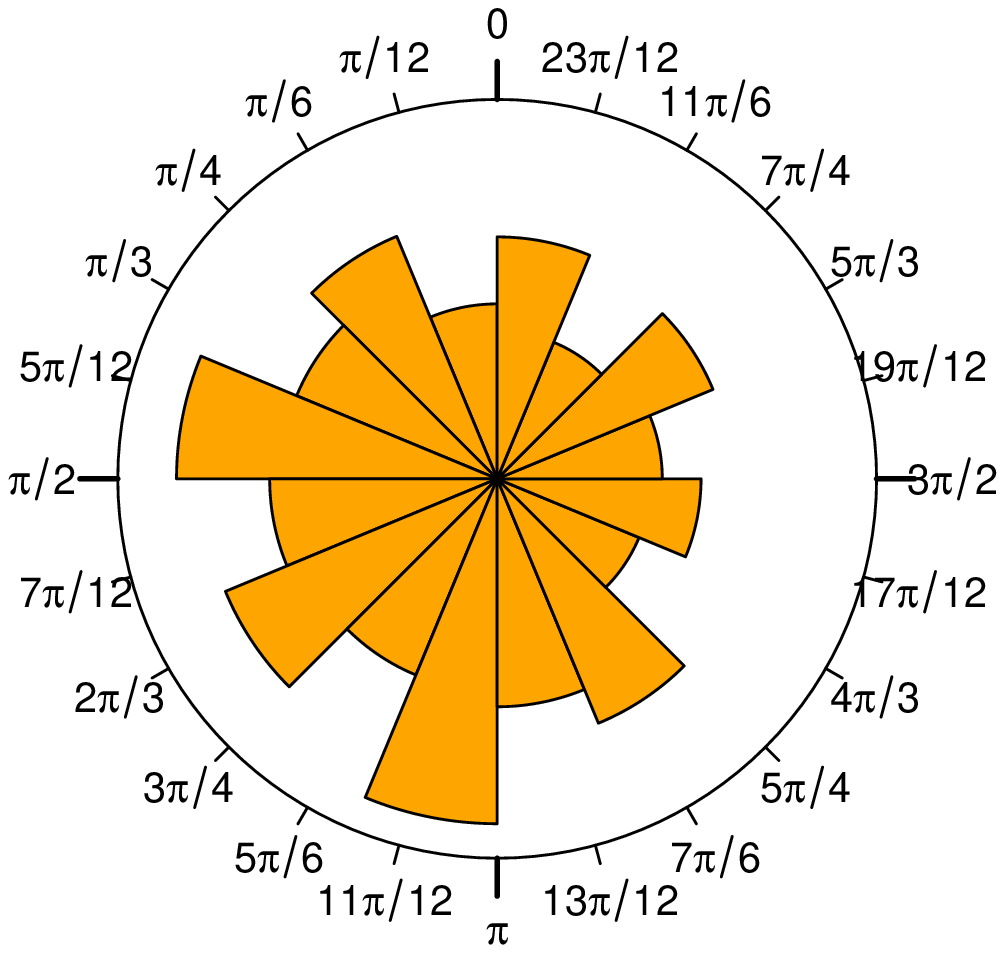}}
   \subfigure[]{\label{rose8}
   \includegraphics[width=0.15\textwidth]{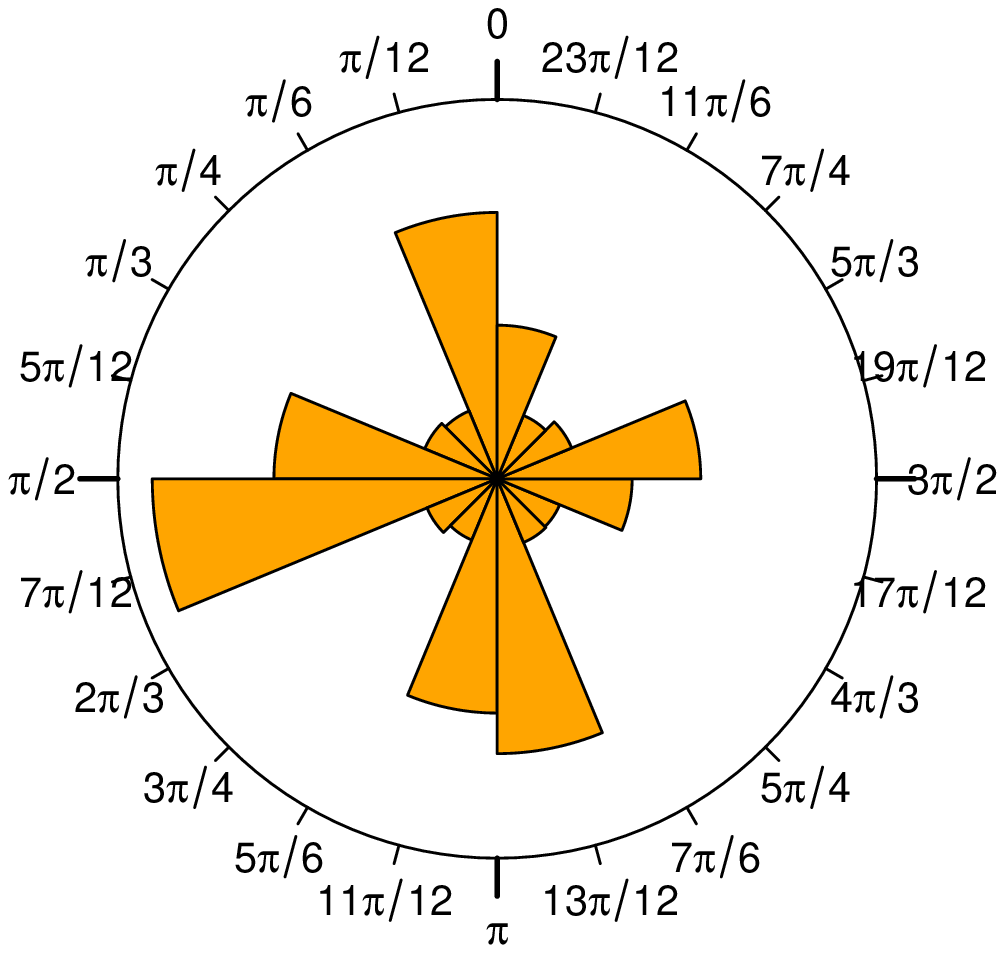}}
    \subfigure[]{\label{rose4}
   \includegraphics[width=0.15\textwidth]{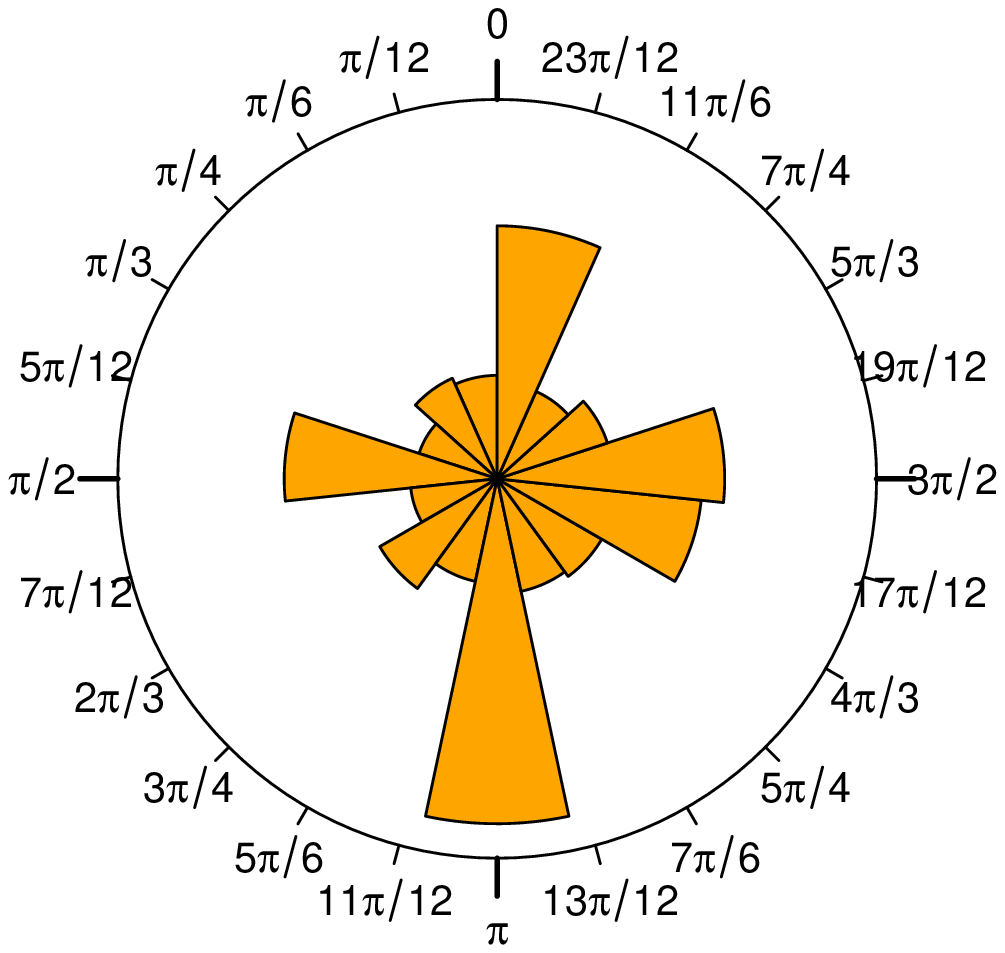}}
\subfigure[]{\label{rose7}
   \includegraphics[width=0.15\textwidth]{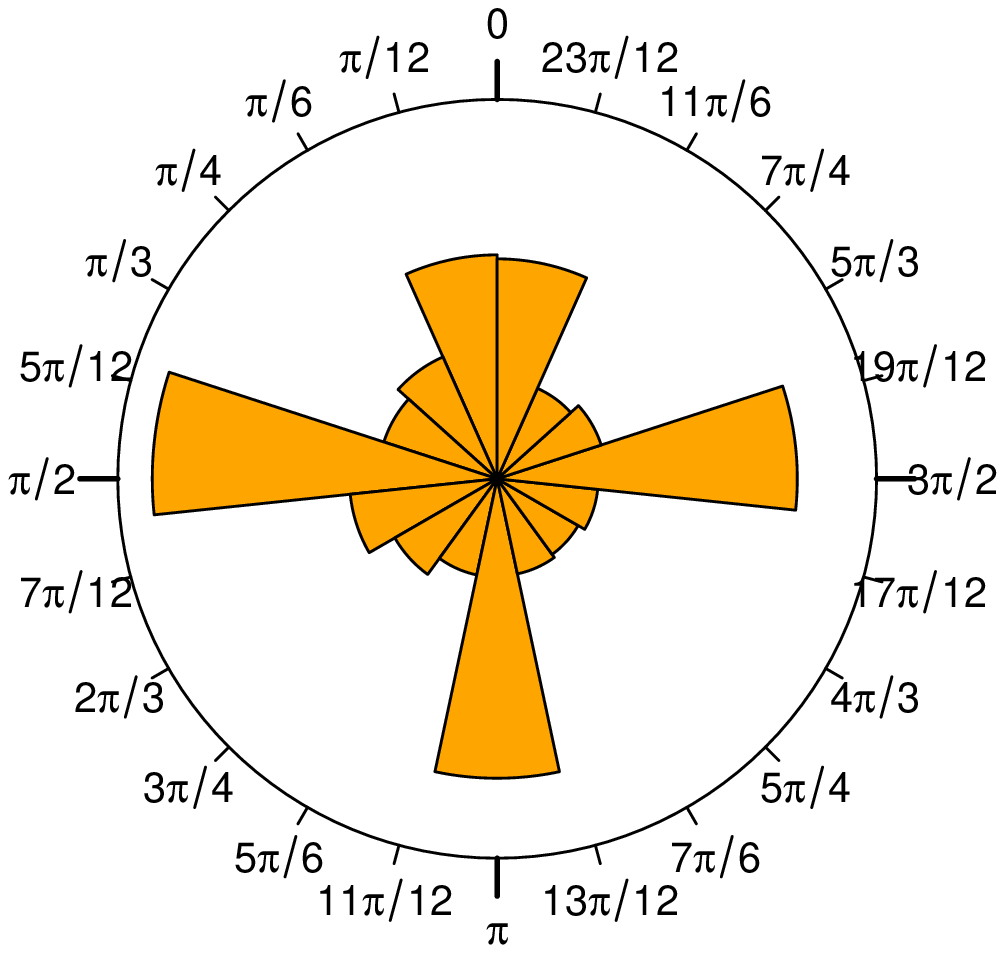}}
   \vspace*{-2mm}     \caption{Beirut's rose diagram is given in \ref{rose1}, Boston's in \ref{rose2}, Boulder in \ref{rose9}, Cambridge's in \ref{rose3}, Lo barnechea's in \ref{rose6}, New York's in \ref{rose5}, Portland's in \ref{rose8}, San Francisco's in \ref{rose4}, and finally Washington's in \ref{rose7}.}      
      \label{mapandrose}
\end{figure}

A building's weighted orientation is defined to be the angle that the major axis of its circumscribing ellipse makes with the North as shown in Fig \ref{ellipsepoly}. More precisely, when the building's long sides' alignments outweigh those of the short sides the orientation is termed weighted compared to its unweighted counterpart where all the sides' directions are equally significant \cite{Mohajeri:2016ev}.
Further, we compute  the streets' orientations with respect to the North using the \textit{maptools}  and \textit{spatstat} packages in \textit{R} \cite{bivand2013maptools}, which allowed us to produce the cities' rose diagrams of in Fig. \ref{mapandrose}, showing the distributions of their streets' orientations. 

Moreover, in order to quantify the dispersal in the orientation we resort to the computation of the entropy \cite{wilson2011entropy,Gudmundsson:2013bk}. 
Particularly, the streets' orientation entropy measures the variability in their respective azimuths and  similarly the buildings' orientation entropy measures the diversity in their major axes alignments. They are respectively given by:

  \begin{equation} \label{streetorientationentropy}
\mathcal{S}_{\mathrm {street \ orientation}}= -\sum_i^{N} p_i \log{p_i}, 
\end{equation}

  \begin{equation} \label{buildingorientationentropy}
\mathcal{S}_{\mathrm {building \ orientation}}= -\sum_i^{N} p_i \log{p_i}, 
\end{equation}
 where $N$ is the number of bins, $p_i$ is the probability that a street or a building is oriented along a direction $i $ with respect to the North with $i$ going from $0$ to $\pi$ in steps of $\pi/12$, and finally $\sum_i^N p_i = 1$. In the case where the distribution is uniform the entropy is $\log{N}$, whereas in the case where the distribution is peaked the entropy is $0$. 
In addition to $\mathcal{S}_{\mathrm {street \ orientation}}$ the network circuitry, defined as the ratio of the sum of all the network's pairwise distances $D$ to the total pairwise Euclidean counterpart $D_e$, characterizes the road network. It is given by:

  \begin{equation} \label{circuitry}
\mathcal{C}= \frac{D}{D_e}, 
\end{equation}
These metrics are calculated using Eq.\ref{streetorientationentropy}-\ref{circuitry} for all the cities and are given in Table \ref{my-label} along with their corresponding city's solar potential $\mathcal{P}$. In what follows we explore their interdependence. 

Figure \ref{scaling1} shows the variation of the buildings orientation entropy as the a function of that of the streets orientation and reveals two scaling regimes given in Eq. \ref{entropies}: 

   \begin{figure}[!htp]%
\includegraphics[width=0.4\textwidth]{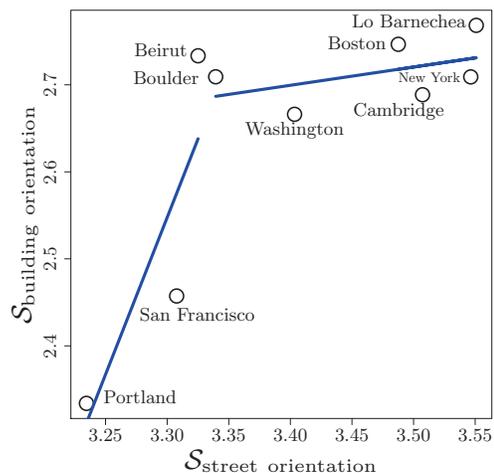}
  \caption{The figure shows the variation in the building entropy as a function of the street orientation entropy. The different regimes are fitted with the blue lines given respectively by: $2.51 + 0.24 \mathcal{S}_{\mathrm {street \ orientation}} $ and $ 2.71 + 0.04 \mathcal{S}_{\mathrm {street \ orientation}}  $ }
  {\label{scaling1}}
\end{figure}

 \begin{equation} \label{entropies}
\frac{\partial \mathcal{S_{\substack {\mathrm {building } \\  {\mathrm {orientation}}}}}}{\partial \mathcal{S_{\substack {\mathrm {street } \\  {\mathrm {orientation}}}}}} = 
\begin{cases}
0.24 \text{ for }\mathcal{S}_{\mathrm {street \ orientation}} \leq 3.32 \\
0.04 \text{ for } \mathcal{S}_{\mathrm {street \ orientation}} > 3.32
\end{cases}
\end{equation}

\begin{figure}[!htp]%
\includegraphics[width=0.4\textwidth]{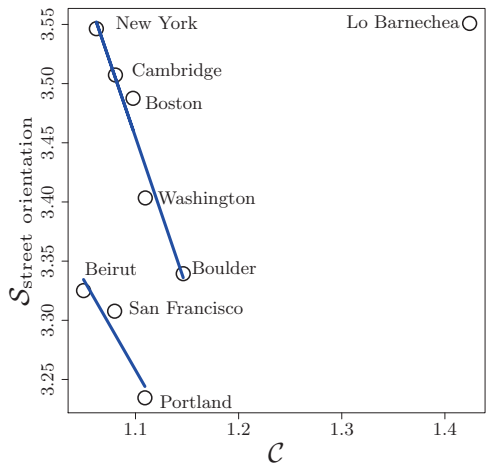}
 \caption{The figure shows the variation in street orientation entropy as a function of circuitry. The different regimes are fitted with the blue lines given respectively by: $ 3.29 -0.06  \ \mathcal{C} $ and $ 3.45 - 0.16 \ \mathcal{C} $ }  {\label{scaling2}}
\end{figure}

Next, we follow the variation in $\mathcal{S}_{\mathrm {street \ orientation}}$ as a function of $\mathcal{C}$, which is shown in Fig. \ref{scaling2}. This also reveals two domains separated by $\mathcal{S}_{\mathrm {street \ orientation}} =3.32$ with their respective slopes given in Eq. \ref{entropycirc}.

 \begin{equation} \label{entropycirc}
\frac{\partial \mathcal{C}}{\partial \mathcal{S}}_{\substack {\mathrm {street } \\  {\mathrm {orientation}}}}= 
\begin{cases}
-0.06 \ \text{for }\mathcal{S}_{\mathrm {street \ orientation}} \leq 3.32 \\
-0.16 \ \text{for } \mathcal{S}_{\mathrm {street \ orientation}} > 3.32
\end{cases}
\end{equation}
 
Subsequently, the network circuitry, the streets' orientation, and the that of the buildings are interdependent, which we expect to be affecting their corresponding solar potential. Therefore, we follow $\mathcal{P}$ as a function of $\mathcal{S}_{\mathrm {street \ orientation}}$, which is shown in Fig. \ref{scaling3}, which in turn exhibits two scaling domains whose' slopes given by Eq. \ref{solarpotentialvsstreetorientationentropy}: 

  \begin{figure}[!htp]%
   \includegraphics[width=0.42\textwidth]{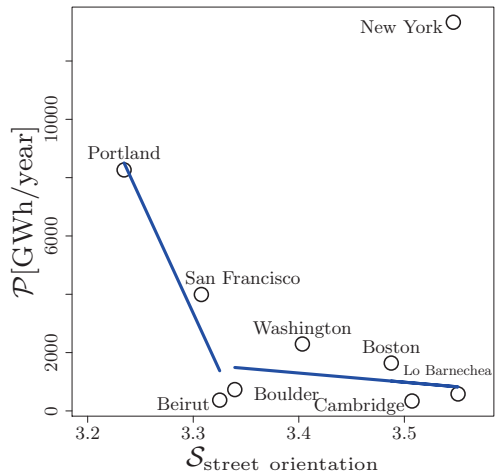}
  \caption{The cities' solar potentials are followed and the blue lines  are given by $-5349 \mathcal{S}_{\mathrm {street \ orientation}} + 4209$, and $ -535 \mathcal{S}_{\mathrm {street \ orientation}} + 1118$.}
  {\label{scaling3}}
\end{figure}
 \begin{equation} \label{solarpotentialvsstreetorientationentropy}
\frac{\partial \mathcal{P}}{\partial \mathcal{S}}_{\substack {\mathrm {street } \\  {\mathrm {orientation}}}} =
\begin{cases}
 \alpha \text{ for } \mathcal{S}_{\mathrm {street \ orientation}} > 3.32 \\
 \beta  \text{ for }  \mathcal{S}_{\mathrm {street \ orientation}} \leq 3.32
\end{cases}
\end{equation}
where $\alpha = -5349$ and $\beta= -535$. 

\begin{figure}[!htp]%
   \includegraphics[width=0.38\textwidth]{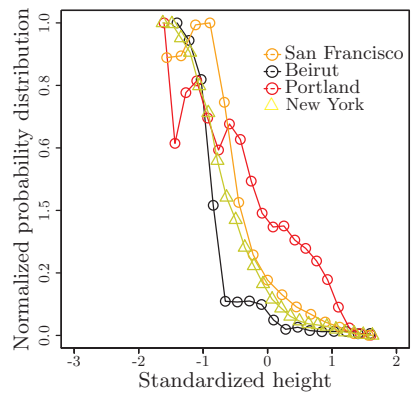}
  \caption{ The figure shows the long-tailed normalized probability distributions of the cities as a function of their respective standardized heights. }
  {\label{dist1}}
\end{figure}

\begin{figure}[!htp]%
   \includegraphics[width=0.35\textwidth]{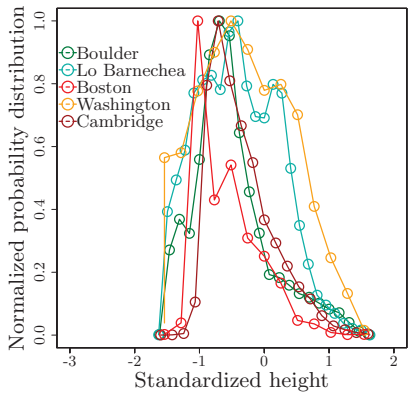}
  \caption{The figure shows the normalized probability distributions of the cities as a function of their respective standardized heights all of which are nearly symmetrical.}
  {\label{dist2}}
\end{figure}

We note that Portland, Beirut, and San Francisco are all hilly cities compared to the others. Thus we suspect that the landform constraints are limiting the streets' orientations as well as those of the buildings and thus this might explain why their respective $\mathcal{S}_{\mathrm {street \ orientation}} $ and $\mathcal{S}_{\mathrm {building \ orientation}} $ are lower compared to their counterparts of Table \ref{my-label}. To explore the effect of landform on these metrics we evaluate the cities' sea-level height distributions calculated using their respective $1.2m$ resolution digital elevations retrieved from Trimble Marketplace \cite{trimble}. Beirut, San Francisco, and Portland's normalized probability distributions of the standardized heights are given in Fig. \ref{dist1}, which appear to be long-tailed distributions in addition to New York's, while those of the rest of the cities are given in Fig. \ref{dist2} and are all nearly symmetrical.
The long-tailed distribution results from the existence of a lower bound on the heights, while the absence thereof brings about symmetrical distributions \cite{Gillespie:2015hb,Mitzenmacher:2004cx,Clauset:2009iy,Bettencourt:2007ej,Newman:2005gv},  appearing respectively in Fig. \ref{dist1} and Fig. \ref{dist2}.

\begin{figure}[!htp]
   \includegraphics[width=0.32\textwidth]{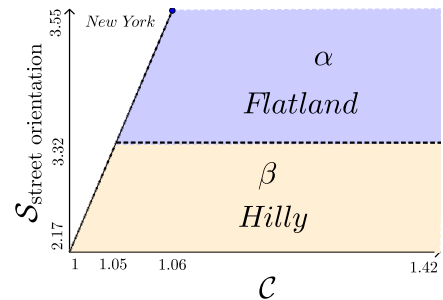}
   \caption{The phase diagram shows the scaling domains $\alpha$ and $\beta$ of $\mathcal{P}$ with $\mathcal{S}_{\mathrm {street \ orientation}}$ , which correspond to  flat and hilly topographies respectively. }
   \label{phasespace}
   \end{figure}

New York's high $\mathcal{S}_{\mathrm {building \ orientation}} $ can be explained by the fact that the city is divided into three major mainlands: Staten Island, Brooklyn and Manhattan each of which has its own orientation as a whole, which is manifested at both the street and building levels despite the fact that the height distribution is long-tailed. 
Moreover, Lo Barnechea's circuitry is higher that the rest of the cities by design. 
Therefore, using the results of Eq. \ref{entropies}-\ref{solarpotentialvsstreetorientationentropy} we can explore the joint effect of $\mathcal{S}_{\mathrm {street \ orientation}}$ and $\mathcal{C}$ and produce the system's phase diagram. We denote the two different regimes of the behavior of $ \mathcal{P}$ as a function of $\mathcal{S}_{\mathrm {street \ orientation}} $ by their corresponding slopes $\alpha$ and $\beta$. The phase diagram is given in Fig. \ref{phasespace}, with New York being the singularity. For values of $\mathcal{S}_{\mathrm {street \ orientation}} \leq 3.32 $, that is for hilly cities, $\mathcal{P}$ scales as $ \beta \mathcal{S}_{\mathrm {street \ orientation}}$ and when $\mathcal{S}_{\mathrm {street \ orientation}} >3.32$, that is for flatlands,  it scales as $\alpha \mathcal{S}_{\mathrm {street \ orientation}}$. This entails that in cities with varying topographies the streets and buildings have a narrow range of orientations along which they are aligned due to the landform constraints, as opposed to flatlands, where they are constructed without restrictions, 
 which can interpreted as two universality classes corresponding to hilly and flat topographies respectively.

It has been shown that landform imposes physical constraints on the alignments of streets \cite{Mohajeri:2014ej,Mohajeri:2014kw} and subsequently on its road network characteristics.  Here we have pushed the idea further and explored the effect of topography in determining the city's buildings' solar potentials as a consequence of the aforementioned constraints. This graph theoretic approach established a clear relation between a city's sustainability and its infrastructure's design subject to landform constraints and complemented our findings which linked $\mathcal{P}$ to its street length distribution \cite{najem2017solar}. 
\bibliography{orientation}

\begin{thebibliography}{26}
\expandafter\ifx\csname natexlab\endcsname\relax\def\natexlab#1{#1}\fi
\expandafter\ifx\csname bibnamefont\endcsname\relax
  \def\bibnamefont#1{#1}\fi
\expandafter\ifx\csname bibfnamefont\endcsname\relax
  \def\bibfnamefont#1{#1}\fi
\expandafter\ifx\csname citenamefont\endcsname\relax
  \def\citenamefont#1{#1}\fi
\expandafter\ifx\csname url\endcsname\relax
  \def\url#1{\texttt{#1}}\fi
\expandafter\ifx\csname urlprefix\endcsname\relax\def\urlprefix{URL }\fi
\providecommand{\bibinfo}[2]{#2}
\providecommand{\eprint}[2][]{\url{#2}}

\bibitem[{\citenamefont{Shepperson}(2017)}]{shepperson2017sunlight}
\bibinfo{author}{\bibfnamefont{M.}~\bibnamefont{Shepperson}},
  \emph{\bibinfo{title}{Sunlight and Shade in the First Cities: A sensory
  archaeology of early Iraq}}, vol.~\bibinfo{volume}{1}
  (\bibinfo{publisher}{Vandenhoeck \& Ruprecht}, \bibinfo{year}{2017}).

\bibitem[{\citenamefont{Paz and Greenberg}(2016)}]{paz2016conceiving}
\bibinfo{author}{\bibfnamefont{S.}~\bibnamefont{Paz}} \bibnamefont{and}
  \bibinfo{author}{\bibfnamefont{R.}~\bibnamefont{Greenberg}},
  \bibinfo{journal}{Journal of Mediterranean Archaeology}
  \textbf{\bibinfo{volume}{29}}, \bibinfo{pages}{197} (\bibinfo{year}{2016}).

\bibitem[{\citenamefont{Mohajeri et~al.}(2016)\citenamefont{Mohajeri, Upadhyay,
  Gudmundsson, Assouline, K{\"a}mpf, and Scartezzini}}]{Mohajeri:2016ev}
\bibinfo{author}{\bibfnamefont{N.}~\bibnamefont{Mohajeri}},
  \bibinfo{author}{\bibfnamefont{G.}~\bibnamefont{Upadhyay}},
  \bibinfo{author}{\bibfnamefont{A.}~\bibnamefont{Gudmundsson}},
  \bibinfo{author}{\bibfnamefont{D.}~\bibnamefont{Assouline}},
  \bibinfo{author}{\bibfnamefont{J.}~\bibnamefont{K{\"a}mpf}},
  \bibnamefont{and} \bibinfo{author}{\bibfnamefont{J.-L.}
  \bibnamefont{Scartezzini}}, \bibinfo{journal}{Renewable Energy}
  \textbf{\bibinfo{volume}{93}}, \bibinfo{pages}{469} (\bibinfo{year}{2016}).

\bibitem[{\citenamefont{Mohajeri and Gudmundsson}(2014)}]{Mohajeri:2014ej}
\bibinfo{author}{\bibfnamefont{N.}~\bibnamefont{Mohajeri}} \bibnamefont{and}
  \bibinfo{author}{\bibfnamefont{A.}~\bibnamefont{Gudmundsson}},
  \bibinfo{journal}{Journal of Geographical Sciences}
  \textbf{\bibinfo{volume}{24}}, \bibinfo{pages}{363} (\bibinfo{year}{2014}).

\bibitem[{\citenamefont{Hillier and Hanson}(1984)}]{hillier1984social}
\bibinfo{author}{\bibfnamefont{B.}~\bibnamefont{Hillier}} \bibnamefont{and}
  \bibinfo{author}{\bibfnamefont{J.}~\bibnamefont{Hanson}},
  \bibinfo{journal}{Cambridge: Press syndicate of the University of Cambridge}
  (\bibinfo{year}{1984}).

\bibitem[{\citenamefont{Bovill}(1996)}]{bovill1996fractal}
\bibinfo{author}{\bibfnamefont{C.}~\bibnamefont{Bovill}},
  \emph{\bibinfo{title}{Fractal geometry in architecture and design}}
  (\bibinfo{publisher}{Springer}, \bibinfo{year}{1996}).

\bibitem[{\citenamefont{Batty and Longley}(1994)}]{batty1994fractal}
\bibinfo{author}{\bibfnamefont{M.}~\bibnamefont{Batty}} \bibnamefont{and}
  \bibinfo{author}{\bibfnamefont{P.~A.} \bibnamefont{Longley}},
  \emph{\bibinfo{title}{Fractal cities: a geometry of form and function}}
  (\bibinfo{publisher}{Academic press}, \bibinfo{year}{1994}).

\bibitem[{\citenamefont{Giacomin and Levinson}(2015)}]{Giacomin:2015iv}
\bibinfo{author}{\bibfnamefont{D.~J.} \bibnamefont{Giacomin}} \bibnamefont{and}
  \bibinfo{author}{\bibfnamefont{D.~M.} \bibnamefont{Levinson}},
  \bibinfo{journal}{Environment and Planning B: Planning and Design}
  \textbf{\bibinfo{volume}{42}}, \bibinfo{pages}{1040} (\bibinfo{year}{2015}).

\bibitem[{\citenamefont{Levinson}(2011)}]{Levinson:2011gc}
\bibinfo{author}{\bibfnamefont{D.}~\bibnamefont{Levinson}},
  \bibinfo{journal}{PLoS ONE} \textbf{\bibinfo{volume}{7}},
  \bibinfo{pages}{e29721} (\bibinfo{year}{2011}).

\bibitem[{\citenamefont{Louf and Barthelemy}(2014)}]{louf2014typology}
\bibinfo{author}{\bibfnamefont{R.}~\bibnamefont{Louf}} \bibnamefont{and}
  \bibinfo{author}{\bibfnamefont{M.}~\bibnamefont{Barthelemy}},
  \bibinfo{journal}{Journal of The Royal Society Interface}
  \textbf{\bibinfo{volume}{11}}, \bibinfo{pages}{20140924}
  (\bibinfo{year}{2014}).

\bibitem[{\citenamefont{Mohajeri et~al.}(2014)\citenamefont{Mohajeri,
  Gudmundsson, and K{\"a}mpf}}]{Mohajeri:2014kw}
\bibinfo{author}{\bibfnamefont{N.}~\bibnamefont{Mohajeri}},
  \bibinfo{author}{\bibfnamefont{A.}~\bibnamefont{Gudmundsson}},
  \bibnamefont{and}
  \bibinfo{author}{\bibfnamefont{J.}~\bibnamefont{K{\"a}mpf}}, in
  \emph{\bibinfo{booktitle}{Proceedings of the EuroGraphics 2014 on Urban Data
  Modelling and Visualisation}} (\bibinfo{year}{2014}).

\bibitem[{\citenamefont{Barth{\'e}lemy}(2010)}]{2011PhR...499....1B}
\bibinfo{author}{\bibfnamefont{M.}~\bibnamefont{Barth{\'e}lemy}},
  \bibinfo{journal}{Physics Reports} \textbf{\bibinfo{volume}{499}},
  \bibinfo{pages}{1} (\bibinfo{year}{2010}).

\bibitem[{\citenamefont{Levinson}(2012)}]{Levinson:2012gca}
\bibinfo{author}{\bibfnamefont{D.}~\bibnamefont{Levinson}},
  \bibinfo{journal}{PLoS ONE} \textbf{\bibinfo{volume}{7}},
  \bibinfo{pages}{e29721} (\bibinfo{year}{2012}).

\bibitem[{\citenamefont{Gudmundsson and Mohajeri}(2013)}]{Gudmundsson:2013bk}
\bibinfo{author}{\bibfnamefont{A.}~\bibnamefont{Gudmundsson}} \bibnamefont{and}
  \bibinfo{author}{\bibfnamefont{N.}~\bibnamefont{Mohajeri}},
  \bibinfo{journal}{Nature Scientific Reports} \textbf{\bibinfo{volume}{3}},
  \bibinfo{pages}{47} (\bibinfo{year}{2013}).

\bibitem[{\citenamefont{Najem}(2017)}]{najem2017solar}
\bibinfo{author}{\bibfnamefont{S.}~\bibnamefont{Najem}},
  \bibinfo{journal}{Physical Review E} \textbf{\bibinfo{volume}{95}},
  \bibinfo{pages}{012323} (\bibinfo{year}{2017}).

\bibitem[{Open Street Map Project()}]{osm2016}
Open Street Map Project (\bibinfo{year}{2016}),
  \bibinfo{note}{www.openstreetmap.org}.

\bibitem[{Mapdwell()}]{mapdwell2016}
Mapdwell (\bibinfo{year}{2016}), \bibinfo{note}{www.mapdwell.com/en/solar}.

\bibitem[{\citenamefont{Fu and Rich}(2000)}]{fu2000solar}
\bibinfo{author}{\bibfnamefont{P.}~\bibnamefont{Fu}} \bibnamefont{and}
  \bibinfo{author}{\bibfnamefont{P.}~\bibnamefont{Rich}},
  \bibinfo{journal}{Helios Environmental Modeling Institute}
  \textbf{\bibinfo{volume}{1616}} (\bibinfo{year}{2000}).

\bibitem[{\citenamefont{Bivand and Lewin-Koh}(2013)}]{bivand2013maptools}
\bibinfo{author}{\bibfnamefont{R.}~\bibnamefont{Bivand}} \bibnamefont{and}
  \bibinfo{author}{\bibfnamefont{N.}~\bibnamefont{Lewin-Koh}},
  \bibinfo{journal}{R package version 0.8-39}  (\bibinfo{year}{2013}),
  \urlprefix\url{https://CRAN.R-project.org/package=maptools}.

\bibitem[{\citenamefont{Wilson}(2011)}]{wilson2011entropy}
\bibinfo{author}{\bibfnamefont{A.~G.} \bibnamefont{Wilson}},
  \emph{\bibinfo{title}{Entropy in urban and regional modelling}},
  vol.~\bibinfo{volume}{1} (\bibinfo{publisher}{Routledge},
  \bibinfo{year}{2011}).

\bibitem[{tri()}]{trimble}
\emph{\bibinfo{title}{Trimble marketplace}},
  \bibinfo{howpublished}{\url{https://market.trimbledata.com}},
  \bibinfo{note}{accessed: 2017-04}.

\bibitem[{\citenamefont{Gillespie}(2015)}]{Gillespie:2015hb}
\bibinfo{author}{\bibfnamefont{C.~S.} \bibnamefont{Gillespie}},
  \bibinfo{journal}{Journal of Statistical Software}
  \textbf{\bibinfo{volume}{64}} (\bibinfo{year}{2015}).

\bibitem[{\citenamefont{Mitzenmacher}(2004)}]{Mitzenmacher:2004cx}
\bibinfo{author}{\bibfnamefont{M.}~\bibnamefont{Mitzenmacher}},
  \bibinfo{journal}{Internet Mathematics} \textbf{\bibinfo{volume}{1}},
  \bibinfo{pages}{226} (\bibinfo{year}{2004}).

\bibitem[{\citenamefont{Clauset et~al.}(2009)\citenamefont{Clauset, Shalizi,
  and Newman}}]{Clauset:2009iy}
\bibinfo{author}{\bibfnamefont{A.}~\bibnamefont{Clauset}},
  \bibinfo{author}{\bibfnamefont{C.~R.} \bibnamefont{Shalizi}},
  \bibnamefont{and} \bibinfo{author}{\bibfnamefont{M.~E.~J.}
  \bibnamefont{Newman}}, \bibinfo{journal}{SIAM Review}
  \textbf{\bibinfo{volume}{51}}, \bibinfo{pages}{661} (\bibinfo{year}{2009}).

\bibitem[{\citenamefont{Bettencourt et~al.}(2007)\citenamefont{Bettencourt,
  Lobo, Helbing, Kuhnert, and West}}]{Bettencourt:2007ej}
\bibinfo{author}{\bibfnamefont{L.~M.~A.} \bibnamefont{Bettencourt}},
  \bibinfo{author}{\bibfnamefont{J.}~\bibnamefont{Lobo}},
  \bibinfo{author}{\bibfnamefont{D.}~\bibnamefont{Helbing}},
  \bibinfo{author}{\bibfnamefont{C.}~\bibnamefont{Kuhnert}}, \bibnamefont{and}
  \bibinfo{author}{\bibfnamefont{G.~B.} \bibnamefont{West}},
  \bibinfo{journal}{Proceedings of the National Academy of Sciences}
  \textbf{\bibinfo{volume}{104}}, \bibinfo{pages}{7301} (\bibinfo{year}{2007}).

\bibitem[{\citenamefont{Newman}(2005)}]{Newman:2005gv}
\bibinfo{author}{\bibfnamefont{M.}~\bibnamefont{Newman}},
  \bibinfo{journal}{Contemporary Physics} \textbf{\bibinfo{volume}{46}},
  \bibinfo{pages}{323} (\bibinfo{year}{2005}).

\end{thebibliography}
\end{document}